\documentstyle[seceq,psfig]{ptptex}

\def\lesssim{\mathrel{\mathchoice {\vcenter{\offinterlineskip\halign{\hfil
$\displaystyle##$\hfil\cr<\cr\sim\cr}}}
{\vcenter{\offinterlineskip\halign{\hfil$\textstyle##$\hfil\cr<\cr\sim\cr}}}
{\vcenter{\offinterlineskip\halign{\hfil$\scriptstyle##$\hfil\cr<\cr\sim\cr}}}
{\vcenter{\offinterlineskip\halign{\hfil$\scriptscriptstyle##$\hfil\cr<\cr\sim\cr}}}}}
\def\grsim{\mathrel{\mathchoice {\vcenter{\offinterlineskip\halign{\hfil
$\displaystyle##$\hfil\cr>\cr\sim\cr}}}
{\vcenter{\offinterlineskip\halign{\hfil$\textstyle##$\hfil\cr>\cr\sim\cr}}}
{\vcenter{\offinterlineskip\halign{\hfil$\scriptstyle##$\hfil\cr>\cr\sim\cr}}}
{\vcenter{\offinterlineskip\halign{\hfil$\scriptscriptstyle##$\hfil\cr>\cr\sim\cr}}}}}

\title{
Gravitational-Wave Driven Instability of Rotating Relativistic Stars 
}

\author{John L.  {\sc Friedman}\footnote{E-mail address: friedman@uwm.edu} and Keith H.  {\sc Lockitch}\footnote{E-mail address: lockitch@gravity.phys.psu.edu}
}

\inst{Department of Physics, \\
University of Wisconsin-Milwaukee,\\ 
PO Box 413, Milwaukee, WI 53201, US\\
and \\
Department of Physics, \\
Pennsylvania State University,\\ 
State College, PA 16802, US}

\recdate{
}

\abst{
A brief review of the stability of rotating relativistic stars
is followed by a more detailed discussion of recent work on an
instability of r-modes, modes of rotating stars that have axial 
parity in the slow-rotation limit.  These modes may dominate the
spin-down of neutron stars that are rapidly rotating at birth, and 
the gravitational waves they emit may be detectable.
}

\begin{document}

\maketitle

\section{Introduction}
A series of recent surprises appear dramatically to have improved 
the likelihood that the spin of rapidly rotating, newly formed neutron 
stars is limited by a nonaxisymmetric instability driven by gravitational
waves -- and that the emitted waves may be detectable.     

The first of these was the discovery that the r-modes,
rotationally restored modes that have axial parity for spherical
models, are unstable in perfect fluid models with arbitrarily slow
rotation.  First indicated in numerical work by Andersson \cite{a97},
the instability is implied in a nearly newtonian context by the
newtonian expression for the r-mode frequency (\ref{p&p_freq}), and a
computation by Friedman and Morsink \cite{jfs97} of the canonical
energy of initial data showed (independent of assumptions on the
existence of discrete modes) that the instability is a generic feature
of axial-parity perturbations of relativistic stars.  

Studies of the viscous and radiative timescales associated with the
r-modes (Lindblom et al.  \cite{lom98}, Owen et al. \cite{o98},
Andersson et al. \cite{aks98}, Kokkotas and Stergioulas \cite{ks98},
Lindblom et al. \cite{lmo99}) have revealed a second surprising
result:  The growth time of r-modes driven by current-multipole
gravitational radiation is significantly shorter than had been
expected, so short, in fact, that the instability to gravitational
radiation reaction easily dominates viscous damping in hot, newly
formed neutron stars (see Fig. \ref{lom} below).  As a result, a
neutron star that is rapidly rotating at birth now appears likely to
spin down by radiating most of its angular momentum in gravitational
waves. (See, however, the caveats below.)

Nearly simultaneous with these theoretical surprises was the discovery
by Marshall et.al. \cite{m98} of a fast (16ms) pulsar in a supernova 
remnant (N157B) in the Large Magellanic Cloud.  Estimates of the initial
period put it in the 6-9ms range, implying the existence of a class of neutron stars that are rapidly rotating at birth. Hence, the newly discovered instability appears to set the upper limit on the spin of the newly discovered class of neutron stars. 

We discuss these developments and the uncertainties that accompany them
below, in the context of a review of the gravitational-wave driven 
instability of rotating stars. The structure and stability of rotating 
relativistic stars has recently been reviewed in detail (Stergioulas 
\cite{s98}, Friedman \cite{f96,f98}, Friedman and Ipser \cite{fi92}), and 
a general discussion of the small oscillations of relativistic stars 
may be found in a recent review article by Kokkotas \cite{k96} 
(see also Kokkotas and Schmidt \cite{ks99}). 

The excitement over the r-mode instability has generated a large 
literature in the past two years.   
(Andersson \cite{a97},
Friedman and Morsink \cite{jfs97},
Kojima \cite{k98},
Lindblom et al. \cite{lom98},
Owen et al. \cite{o98},
Andersson, Kokkotas and Schutz \cite{aks98},
Kokkotas and Stergioulas \cite{ks98},
Andersson, Kokkotas and Stergioulas \cite{akst98},
Madsen \cite{mad98},
Hiscock \cite{h98},
Lindblom and Ipser \cite{li98},
Bildsten \cite{b98},
Levin \cite{l98},
Ferrari et al. \cite{fms98},
Spruit \cite{s99},
Brady and Creighton \cite{bc98},
Lockitch and Friedman \cite{lf98},
Lindblom et al. \cite{lmo99}, 
Beyer and Kokkotas \cite{bk99}, 
Kojima and Hosonuma \cite{kh99}, 
Lindblom \cite{l99}, 
Schneider et al. \cite{sfm99},
Rezzolla et al. \cite{rea99}, 
Yoshida and Lee \cite{yl99})
[Earlier studies of the axial-parity oscillations of 
models of the neutron star crust  were reported by van Horn \cite{vh80} 
and by Schumaker and  Thorne \cite{st83}); and Chandrasekhar and Ferrari 
discussed the resonant scattering of axial wave modes \cite{cf91b} and the 
coupling between axial and polar modes induced by stellar rotation 
\cite{cf91a}.] Although we mention many of the recent contributions listed
here, our summary is too short to allow much detail.

\section{Gravitational-Wave Driven Instability of Rotating Relativistic Stars}

Only recently have the oscillation modes of rotating 
relativistic stars begun to be accessible to numerical 
study (see below). Early work on the perturbations of such stars 
focused mainly on the criteria for their stability, and led to the 
discovery that all rotating perfect fluid stars are 
subject to a nonaxisymmetric instability driven by gravitational 
radiation.  The instability was found by Chandrasekhar 
\cite{ch70} for the $l=m=2$ polar mode of the uniform-density, 
uniformly rotating Maclaurin spheroids.  Although this mode is 
unstable only for rapidly rotating models, by looking at the 
canonical energy of initial data with arbitrary values of $m$, Friedman and 
Schutz \cite{fs78b} and Friedman \cite{f78} showed that the instability 
is a generic feature of rotating perfect fluid stars, that 
even slowly rotating perfect-fluid models are formally unstable.  

For a normal mode of the form 
$e^{i(\sigma t+m\varphi)}$ this nonaxisymmetric instability 
acts in the following manner:  
In a non-rotating star, gravitational radiation removes positive 
angular momentum from a forward moving mode and negative angular 
momentum from a backward moving mode, thereby damping all 
time-dependent, non-axisymmetric modes. In a star rotating 
sufficiently fast, however, a backward moving mode can be dragged 
forward as seen by an inertial observer; and it will then radiate
positive angular momentum.  The angular momentum of the mode, however,
remains negative, because the perturbed star has lower total angular 
momentum than the unperturbed star.  As positive angular momentum
is removed from a mode with negative angular momentum, the angular 
momentum of the mode becomes increasingly negative, implying that
its amplitude increases: The mode is driven by gravitational radiation.

The conclusion, that a mode is unstable if it is prograde relative to 
infinity and retrograde relative to the star is equivalent to requiring
that its frequency satisfies the condition,
\begin{equation}
\sigma(\sigma+m\Omega) < 0.
\label{criterion}
\end{equation}
For the polar f- and p-modes, the frequency is large and 
approximately real.  Condition (\ref{criterion}) will be met 
only if $|m\Omega|$ is of order $|\sigma|$, so that for a given 
angular velocity the instability will set in first through modes 
with large $m$. 

\begin{figure}
\centerline{\psfig{file=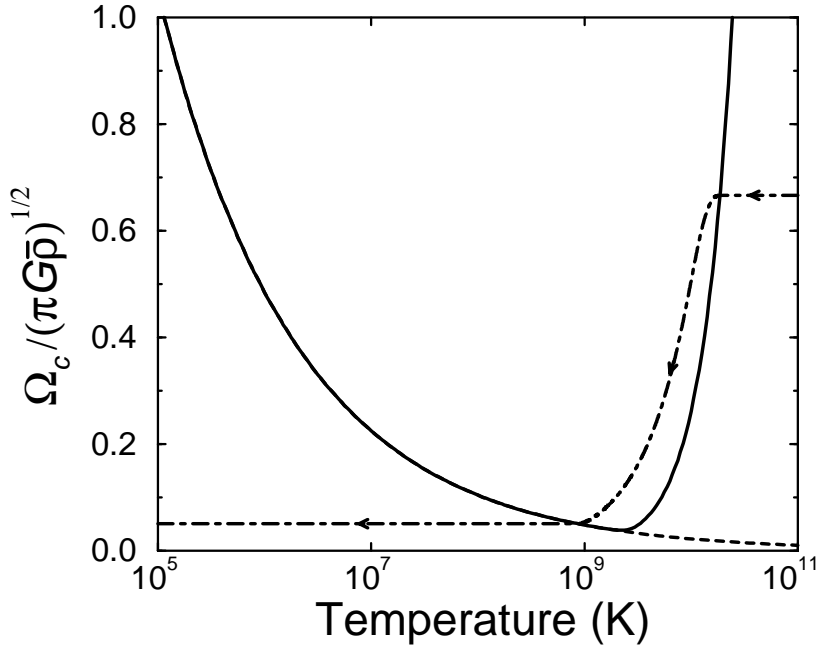,height=3.5in}}
\caption{Critical angular velocity vs. temperature for an $n=1.0$ 
polytrope.  Above the solid curve, the star rotates rapidly enough for its
fastest growing ($l=m=2$) r-mode to be unstable, whereas below the
curve all modes are damped by viscosity.  The dashed curve does not 
include the effects of bulk viscosity, while the solid curve does; 
however, for more accurate calulations of the bulk viscosity contribution 
see Lindblom, Mendell and Owen \cite{lmo99} and Andersson, Kokkotas and Schutz 
\cite{aks98}.  The dot-dashed curve shows the evolution of a rapidly rotating 
neutron star as it cools and spins down due to the emission of gravitational 
waves. [This figure is due to Lindblom, Owen and Morsink \cite{lom98} 
and is reproduced here by permission of the authors.]}
\label{lom}
\end{figure}

The instability spins a star down by allowing it to radiate
away its angular momentum in gravitational waves.  However, to
determine whether this mechanism may be responsible for limiting
the rotation rates of actual neutron stars, one must also 
consider the effects of viscous damping on the perturbations.  
Detweiler and Lindblom \cite{dl77} suggested that viscosity would
stabilize any mode whose growth time was longer than the viscous
damping time, and this was confirmed by Lindblom and Hiscock 
\cite{lh83}. Recent work has indicated that the 
gravitational-wave-driven instability can only limit the rotation 
rate of hot neutron stars, with temperatures above the superfluid 
transition point, $T\sim 10^9\mbox{K}$, but below the temperature 
at which bulk viscosity apparently damps all modes, 
$T\sim 10^{10}\mbox{K}$ (see Fig. \ref{lom}). 
(Ipser and Lindblom \cite{il91}; Lindblom 
\cite{l95} and Lindblom and Mendell \cite{lm95}) 
Because of uncertainties in the temperature of the superfluid phase 
transition and in our understanding of the dominant mechanisms for 
effective viscosity, even this brief temperature window is not 
guaranteed.

The instability appears to arise in modes with both polar and axial
parity, with the distinction defined as follows.  The spherical
symmetry of a nonrotating star implies that its perturbations can be divided
into two classes, polar and axial, according to their behavior under
parity. Where polar tensor fields on a 2-sphere can be constructed from
the scalars $Y_l^m$ and their gradients $\nabla Y_l^m$ (and the metric
on a 2-sphere), axial fields involve the pseudo-vector $\hat r\times
\nabla Y_l^m$, and their behavior under parity is opposite to that of
$Y_l^m$. That is, axial perturbations of odd $l$ are invariant under
parity, and axial perturbations with even $l$ change sign.  Because a
rotating star is also invariant under parity, its perturbations can
also be classified according to their behavior under parity.  Although
$l$ is well-defined only  for modes of the spherical configuration, if
a mode varies continuously along a sequence of equilibrium
configurations that starts with a spherical star and continues along a
path of increasing rotation, the mode can be called axial (polar) if it is
axial (polar) for the spherical star.
 
It is useful to subdivide stellar perturbations according to the
physics dominating their behaviour.  This classification was first
developed by Cowling \cite{c41} for the polar perturbations of
newtonian polytropic models.  The p-modes are polar-parity modes having
pressure as their dominant restoring force \footnote{The lowest p-mode
for each value of $l$ and $m$ is termed an f-mode or fundamental
mode.}  They typically have large pressure and density perturbations
and high frequencies (higher than a few kilohertz for neutron stars).
The other class of polar-parity modes are the g-modes, which are
chiefly restored by gravity.  They typically have very small pressure
and density perturbations and low frequencies.  Indeed, for isentropic
stars, which are marginally stable to convection, the g-modes are all
zero-frequency and have vanishing perturbed pressure and density.  
Similarly, all axial-parity perturbations of newtonian
perfect fluid models have zero frequency in a non-rotating star. The
perturbed pressure and density as well as the radial component of the
fluid velocity are all rotational scalars and must have polar parity.
Thus, the axial perturbations of a spherical star are simply stationary
horizontal fluid currents.

The analogues of these modes in relativistic models of neutron stars
have been studied by many authors.  More recently, an additional class 
of outgoing modes has been identified that exist only in relativistic 
stars.  Like the modes of black holes, these are essentially associated 
with the dynamical spacetime geometry and have been termed w-modes, or 
gravitational wave modes.  Their existence was first argued by Kokkotas 
and Schutz \cite{ks86}. The polar w-modes were first found by Kojima 
\cite{k88} as rapidly damped modes of weakly relativistic models, while 
the axial w-modes were first studied by Chandrasekhar and Ferrari 
\cite{cf91b} as scattering resonances of highly relativistic models.  
(See the reviews by Kokkotas \cite{k96} and Kokkotas and Schmidt 
\cite{ks99}.)  All w-modes appear to be stable, and if they are 
analogous to the modes of non-interacting zero-rest-mass test fields
on the geometry of a rotating star, they should be unstable only when
an ergosphere is present; but we are not aware of a careful study 
to rule out w-mode instability.

Until recently, the polar p-modes were expected to dominate the CFS
instability through their coupling to mass multipole radiation.  The
relativistic stability points of p-modes have recently been found by
Stergioulas and Friedman \cite{s96,sf97} and, in the Cowling
approximation, by Yoshida and Eriguchi \cite{ye97}.  This last reference also provides the
first computation of the modes of rapidly rotating relativistic stars.
The relativistic computation shows an instability of each at
significantly smaller values of dimensionless measures of rotation than
is seen in nearly newtonian stars.  Particularly striking is the fact
that the l=m=2 bar mode is unstable for relativistic polytropes of
index n=1.0. The classical Newtonian result for the onset of the bar
mode instability, $n<0.808$) is replaced by $n < 1.3$ for relativistic
polytropes.

\begin{figure}
\centerline{\psfig{file=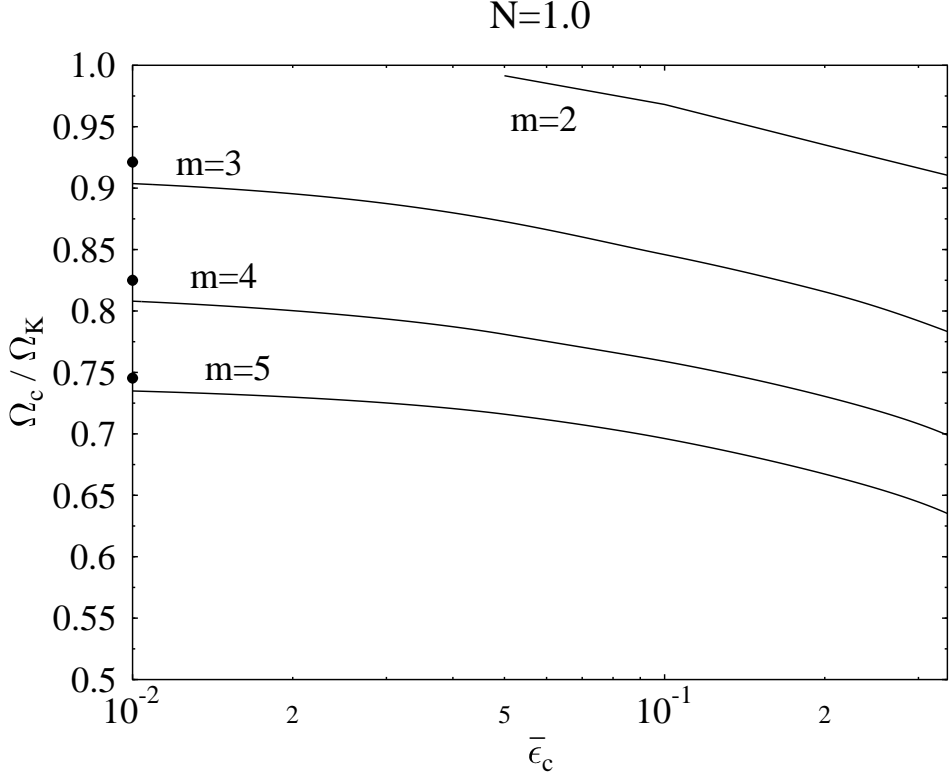,height=5in}}
\caption{Critical ratio of angular velocity to Kepler velocity vs. a 
dimensionless central energy density $\bar{\epsilon_c}$ for the 
$m=2,3,4 \ \mbox{and} \ 5$ neutral modes of $n=1.0$ polytropes.  
The largest value of $\bar{\epsilon_c}$ shown corresponds to the most 
relativistic stable configurations, while the lowest $\bar{\epsilon_c}$ 
corresponds to less relativistic configurations.  The filled circles on 
the vertical axis represent the Newtonian limit.}
\label{omega1.0}
\end{figure}

Fig. \ref{omega1.0} shows the critical angular velocity above which the 
lowest $l=m$ p-modes are unstable in perfect fluid models.  
For most realistic equations of state, Morsink, Stergioulas and Blattnig \cite{msb} 
find that the $l=m=2$ bar mode is unstable in $1.4 M_\odot$ neutron
stars for angular velocities above 0.8$\Omega_K$ to 0.95$\Omega_K$.

The competition between viscosity and gravitational radiation is 
described by the dissipation equation, 

\begin{equation}  \frac1\tau =  \frac1{\tau_{\rm GR}} + 
\frac1{\tau_{\rm shear viscosity}} +\frac1{\tau_{\rm bulk viscosity}},
\end{equation}
with $\tau$ the e-folding time for each process. (The analysis 
sketched below can be found in Lindblom et al\cite{lom98}; see also
Ipser and Lindblom\cite{il91}). When the energy radiated per cycle 
is small compared to the energy 
of the mode, the imaginary part of the mode frequency is accurately 
approximated by the expression
\begin{equation}\frac{1}{\tau} = - \frac{1}{2E} \frac{dE}{dt},   
\label{tau}
\end{equation}
where $E$ is the energy of the mode as measured in the rotating frame,
\begin{equation}E = \frac{1}{2} \int \left[ 
\rho \delta v^a \delta v^{\ast}_a 
+ \left( \frac{\delta p}{\rho}+\delta\Phi\right) 
\delta\rho^{\ast}
\right] d^3 x .
\label{E}
\end{equation}
We have,
\begin{eqnarray}
\frac{dE}{dt} &=& -\sigma (\sigma + m\Omega) \sum_{l\geq 2} N_l \sigma^{2l} \left(
\left|\delta D_{lm}\right|^2   + \left|\delta J_{lm}\right|^2
\right) \nonumber \\
 & & - \int \left( 
2\eta\delta\sigma^{ab}\delta\sigma_{ab}^{\ast} 
+\zeta \delta\theta\delta\theta^{\ast}
\right), 
\label{dEdt}
\end{eqnarray}
where the dissipation due to gravitational radiation \cite{th80} has coupling constant
\begin{equation}N_l = \frac{4\pi G}{c^{2l+1}}\frac{(l+1)(l+2)}{l(l-1)[(2l+1)!!]^2};
\end{equation}
$\delta\sigma_{ab}$ and $\delta\theta$ are the coefficients of shear
and bulk viscosity; and the corresponding coefficients $\eta$ and 
$\zeta$ are estimated (Cutler and Lindblom \cite{cl87}; Sawyer \cite{s89}) by 
\begin{equation}\eta = 2\times 10^{18} 
\left(\frac{\rho}{10^{15}\mbox{g}\!\cdot\!\mbox{cm}^{-3}}\right)^{\frac{9}{4}}
\left(\frac{10^9K}{T}\right)^2 \ 
\mbox{g}\!\cdot\!\mbox{cm}^{-1}\!\cdot\!\mbox{s}^{-1},
\label{eta}
\end{equation}
and 
\begin{equation}
\zeta = 6\times 10^{25} 
\left(\frac{1\mbox{Hz}}{\sigma + m\Omega}\right)^2
\left(\frac{\rho}{10^{15}\mbox{g}\!\cdot\!\mbox{cm}^{-3}}\right)^2
\left(\frac{T}{10^9K}\right)^6 \ 
\mbox{g}\!\cdot\!\mbox{cm}^{-1}\!\cdot\!\mbox{s}^{-1};
\label{zeta}
\end{equation}

Polar and axial radiation arise, respectively, from mass and current
multipoles, $D_{lm}$ and $J_{lm}$, given by the equations,
\begin{equation}
  D_{lm}  = \int   dV r^l  \rho   Y_{lm}^*\qquad\qquad   \ \ \ J_{lm}  = \int  dV r^l  \rho { \bf v  \cdot r} \times\nabla Y_{lm}^*
\end{equation}
The additional factor of $v$ in the current multipoles implies an
additional factor of $v^2$ in the radiated energy of axial modes, 
and hence a smaller expected rate of radiation for the same multipole: 
For a mode of amplitude $A$ = (displacement of fluid element)/$R$,
with $R$ the stellar radius, we have  
\begin{equation}
  \frac{dE}{dt}  \sim A^2   M^2 R^{2l} \sigma^{2l+2}\qquad\qquad 
  \frac{dE}{dt}  \sim A^2   M^2 R^{2l +2} \sigma^{2l +4}. 
\end{equation}

The extra factor of $\sigma^2$ and the fact that $\sigma$ is
proportional to $\Omega$ for slowly rotating stars led to the
incorrect expectation that $r$-modes could be neglected.

\section{The r-mode Instability and its Implications}

The unexpected apparent dominance of the r-modes arises from the fact
that, where the polar $l=2$ mode is unstable, if at all, only for
$\Omega\sim\Omega_K$, the $l=m=2$ r-mode is unstable (for perfect-fluid
models) for arbitrarily small $\Omega$,\cite{a97,jfs97} The instability
in models of slowly rotating, nearly newtonian stars, follows from the
fact that the frequencies, \cite{pp78}
\begin{equation}
\sigma+m\Omega = \frac{2m\Omega}{l(l+1)}.
\label{p&p_freq}
\end{equation}
satisfy the criterion (\ref{criterion}), 
\begin{equation}
\sigma(\sigma+m\Omega) = 
- \frac{2(l-1)(l+2)m^2\Omega^2}{l^2(l+1)^2} < 0.
\end{equation}
[A computation by Friedman and Morsink \cite{jfs97} of the canonical 
energy of initial data showed (independent of assumptions on the existence
of discrete modes) that the instability is a generic feature of 
axial-parity fluid perturbations of relativistic stars.]

Not only are p-modes with low values of $m$ unstable only for large
values of $\Omega$, their frequencies are near zero, because the
frequency of a mode vanishes at its instability point, and a star
cannot rotate fast enough for the lowest p-modes to be far from their
instability points.  Thus where p-modes may limit the rotation of a 
newly formed star to $0.9\Omega_K$,\cite{lm95} 
\footnote{this is likely to decrease, perhaps to 0.8
$\Omega_K$ when the correct general-relativistic computation with
viscosity is completed.}  
the r-modes are likely to limit rotation to less than 0.2$\Omega_K$.
Here $\Omega_K$ is the maximum angular velocity of a uniformly rotating
star, equal to the angular velocity of a satellite in orbit at the
star's equator.

The current picture that has emerged of the spin-down of a hot, 
newly formed neutron star can be readily understood in terms of 
a model of the r-mode instability due to Owen, Lindblom, Cutler, 
Schutz, Vecchio and Andersson (hereafter OLCSVA) \cite{o98}.  
Since one particular mode (with spherical harmonic indices $l=m=2$ 
and frequency $\sigma = -4\Omega/3$) is found to dominate the 
r-mode instability, the perturbed star is treated as a simple system 
with two degrees of freedom: the uniform angular velocity $\Omega$ of 
the equilibrium star, and the (dimensionless) amplitude $\alpha$ of 
the $l=m=2$ r-mode. 
Initially, the neutron star forms with a temperature large 
enough for bulk viscosity to damp any unstable modes, 
$T \grsim 10^{10}\mbox{K}$; the star is assumed to be rotating 
close to its maximum (Kepler) velocity, $\Omega_K\sim \sqrt{M/R^3}$. 
The star then cools by neutrino emission at a rate given by a 
standard power law cooling formula (Shapiro and Teukolsky 
\cite{st83b}). Once it reaches the 
temperature window at which the $l=m=2$ r-mode can go unstable, 
the system is assumed to evolve in three stages.  

First, the amplitude of the r-mode undergoes rapid exponential 
growth from some arbitrary tiny magnitude. Using conservation of 
energy, (\ref{dEdt}) and the related equation for the 
rate of angular momentum loss to gravitational waves, 
\begin{equation}
\frac{dJ}{dt} = 
     -\frac{c^3}{16\pi G}\left(\frac{4\Omega}{3}\right)^5 J_{22}^2,
\end{equation}
with both $E$ and $J$ proportional to $\alpha^2$, 
OLCSVA derive the following equations 
for the evolution of the system in this stage. After 
expressing $E$ and $J$ in terms of $\Omega$ and $\alpha$ 
\begin{equation}
\frac{d\Omega}{dt} = -\frac{2\Omega}{\tau_V}
\frac{\alpha^2 Q}{1+\alpha^2 Q}
\label{owen1}
\end{equation}
\begin{equation}
\frac{d\alpha}{dt} = \frac{\alpha}{|\tau_{GR}|}
-\frac{\alpha}{\tau_V} \frac{1-\alpha^2 Q}{1+\alpha^2 Q}
\label{owen2}
\end{equation}
Here, $\tau_{GR}$ and $\tau_V$ are, respectively, the 
timescales for the growth of the mode by gravitational radiation 
reaction and the damping of the mode by viscosity (see Sect. 2.6).  
(The parameter $Q$ is a constant of order 0.1 related to the 
initial angular
momentum and moment of inertia of the equilibrium star.)
Since the initial amplitude $\alpha$ of the mode is so small, 
the angular momentum changes very little at first 
(Eq. (\ref{owen1})).  That this stage is characterized by
the rapid exponential growth of $\alpha$ is the statement
that the first term in Eq. (\ref{owen2}) (the radiation 
reaction term) dominates over the second (viscous damping).

Eventually the mode will grow to a size at which linear
perturbation theory is insufficient to describe its behavior.
It is expected that a non-linear saturation will occur, halting
the growth of the mode at some amplitude of order unity, 
although the details of these non-linear effects are poorly 
understood at present.  When this saturation occurs, the system 
enters a second evolutionary 
stage during which the mode amplitude remains essentially 
unchanged and the angular momentum of the star is radiated 
away.  During this stage OLCSVA evolve their model system 
according to the equations
\begin{equation}
\alpha^2=\kappa
\label{owen3}
\end{equation}
\begin{equation}
\frac{d\Omega}{dt} = - \frac{2\Omega}{|\tau_{GR}|}
\frac{\kappa Q}{1-\kappa Q}
\label{owen4}
\end{equation}
where $\kappa$ is constant of order unity parameterizing 
the uncertainty in the degree of non-linear saturation.
The star spins down by Eq. (\ref{owen4}), radiating away most 
of its angular momentum while continuing to cool gradually.  

When its temperature and angular velocity are low enough that 
viscosity again dominates the gravitational-wave-driven 
instability, the mode will be damped.  During this third
stage, OLCSVA return to Eqs. (\ref{owen1})-(\ref{owen2}) to 
continue the evolution of their system.  That the mode amplitude 
decays is the statement that the second term in Eq. (\ref{owen2})
(the viscous damping term) dominates the first (radiation
reaction), at this temperature and angular velocity.

This three-stage evolutionary process leaves the newly formed neutron star 
with an angular
velocity small compared with $\Omega_K$.  This final 
angular velocity appears to be fairly insensitive to the 
initial amplitude of the mode and to its degree of non-linear 
saturation. A final period $P\grsim5-10\mbox{ms}$ apparently 
rules out accretion-induced collapse of white dwarfs as a 
mechanism for the formation of millisecond pulsars with 
$P\lesssim 3\mbox{ms}$. 

\begin{figure}
\centerline{\psfig{file=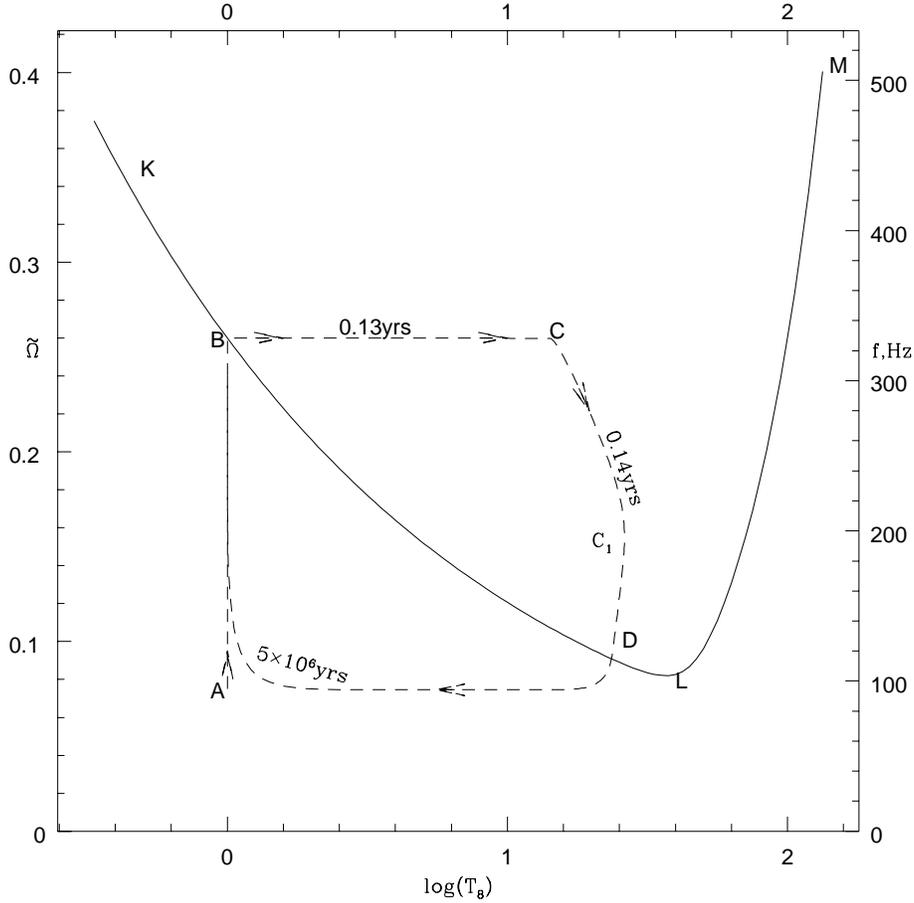,height=5in}}
\caption{Levin's \cite{l98} cyclic evolution scenario for an accreting 
neutron star when r-mode damping is a decreasing function of temperature. 
The solid curve K-L-M shows the critical angular velocity vs. temperature, 
while the dashed curve A-B-C-D-A shows the evolution of the star.
[This figure is due to Levin \cite{l98} and is reproduced 
here by permission of the author.]}
\label{levin}
\end{figure}

The r-mode instability has also revived interest in the Wagoner 
\cite{w84} mechanism, in which old neutron stars are spun up by 
accretion until  the angular momentum loss in gravitational radiation
balances the accretion torque.
Bildsten \cite{b98} and Andersson, Kokkotas and Stergioulas 
\cite{akst98} have proposed that the r-mode instability might 
succeed in this regard where the instability to polar modes 
seems to fail.  However, the mechanism appears to be highly
sensitive to the temperature dependence of viscous damping.
Levin \cite{l98} has argued that if the r-mode damping is a 
decreasing function of temperature (at the temperatures expected
for accreting neutron stars, $T\sim 10^8\mbox{K}$) then viscous
reheating of the unstable neutron star could drive the system 
away from the Wagoner equilibrium state.  Instead, the star would 
follow the cyclic evolution pattern depicted in Fig. \ref{levin}.  
Initially, the runaway reheating would drive the star further into the 
r-mode instability regime (B-C in Fig. \ref{levin}) and 
spin it down to a fraction of its angular velocity (C-D). Once it has 
slowed to the point at which the r-modes become damped, it would 
again slowly cool (D-A) and begin to spin up by accretion (A-B).  Eventually, 
it would again reach the critical angular velocity for the onset of 
instability and repeat the cycle.  Since the radiation
spin-down time is of order 1 year, while the accretion spin-up time 
is of order $10^6$ years, the star spends only a small fraction of the
cycle emitting gravitational waves via the unstable r-modes.  
This would significantly reduce the likelihood that detectable 
gravitational radiation is produced by such sources. 
On the other hand, if the r-mode damping is independent of - or
increases with - temperature (at $T\sim 10^8\mbox{K}$) 
then the Wagoner equilibrium 
state may be allowed (Levin \cite{l98}).  Work is currently in 
progress (Lindblom and Mendell \cite{lm99}) to investigate the r-mode 
damping by mutual friction in superfluid neutron stars, which was the 
dominant viscous mechanism responsible for ruling out the Wagoner 
scenario in the first place (Lindblom and Mendell \cite{lm95}).

Other uncertainties in the scenarios described above are still 
to be investigated.  There is substantial uncertainty in the 
cooling rate of neutron stars, with rapid cooling expected if stars 
have a quark interior or core, or a kaon or pion condensate. Madsen 
\cite{mad98} suggests that an observation of a young neutron star with 
a rotation period below $5-10\mbox{ms}$ would be evidence for a quark 
interior; but even without rapid cooling, the uncertainty in the 
superfluid transition temperature may allow a superfluid to form at 
about $10^{10} \mbox{K}$, possibly killing the instability. 
 
We noted above the expectation that the growth of the unstable 
r-modes will saturate at an amplitude of order unity due to 
non-linear effects (such as mode-mode couplings); however,
this limiting amplitude is not yet known with any certainty and
could be much smaller.  In particular, it has been suggested that 
the non-linear evolution of
the r-modes will wind up the magnetic field of a neutron star, 
draining energy away from the mode and eventually suppressing the
unstable modes entirely (Rezzolla, Lamb and Shapiro \cite{rls99};
see also Spruit \cite{s99}). 

If the r-modes saturate near unity, a study by Brady and
Creighton\cite{bc98} following the work by Owen et al\cite{o98}, finds
that newly formed neutron stars should be detectable by LIGO II with
narrow banding out to about 8 Mpc, with
uncertainty allowing a range of perhaps 4-20 Mpc.  The Virgo cluster 
is then likely to be out of reach, and even with this optimistic assumption
about the saturation amplitude, r-modes are most likely not to be 
detected until sensitivity passes the limits of advanced LIGO.
  
Despite the recent intense interest in r-modes, they are not yet 
well-understood for stellar models appropriate to neutron
stars.  A neutron star is accurately described by a perfect
fluid model in which both the equilibrium and perturbed 
configurations obey the same one-parameter equation 
of state.  We call such models isentropic, 
because isentropic models and their adiabatic perturbations 
obey the same one-parameter equation of state.

For stars with more general equations of state, the r-modes 
describe the dynamical evolution of initial perturbations that 
have axial parity. This is not, however, the case for isentropic 
models. Early work on the r-modes focused on newtonian models 
with general equations 
of state (Papalouizou and Pringle \cite{pp78}, Provost et al. 
\cite{pea81}, Saio \cite{s82}, Smeyers and Martens \cite{sm83})
and mentioned only in passing
the isentropic case.  In isentropic newtonian stars, one finds 
that the only purely axial modes allowed are the r-modes with 
$l=m$ and simplest radial behavior.
(Provost et al. \cite{pea81}\footnote{An appendix in this paper 
incorrectly claims that no $l=m$ r-modes exist, based on an 
incorrect assumption about their radial behavior.})
It is these r-modes only that have been 
studied (and found to be physically interesting) in connection 
with the gravitational-wave driven instability.  

The disappearance of the purely axial modes with $l>m$ occurs for the
following reason.(Lockitch and Friedman \cite{lf98})  Axial
perturbations of a spherical star are time-independent convective
currents with vanishing perturbed pressure and density.  In spherical
{\sl isentropic} stars, stars for which both star and perturbation are
governed by a single one-parameter equation of state, the gravitational
restoring forces that give rise to the g-modes vanish and they, too,
become time-independent convective currents with vanishing perturbed
pressure and density.  Thus, the space of zero frequency modes, which
generally consists only of the axial r-modes, expands for spherical
isentropic stars to include the polar g-modes.  This large degenerate
subspace of zero-frequency modes is split by rotation to zeroth order
in the star's angular velocity, and the corresponding modes of rotating
isentropic stars are generically hybrids whose spherical limits are
mixtures of axial and polar perturbations.

These hybrid rotational modes have already been found analytically for
the uniform-density Maclaurin spheroids by Lindblom and Ipser
\cite{li98} and numerically for slowly rotating $n=1$ polytropes by
Lockitch and Friedman \cite{lf98}; and Yoshida and Lee \cite{yl99} have
obtained these Newtonian polytropic modes to the next nonvanishing
order in $\Omega$. Because the restoring
force for these modes is rotation, as it is for the axial r-modes,
Ipser and Lindblom \cite{li98} call them rotational modes or generalized r-modes.

The r-modes of rotating relativistic stars were studied for the first
time only recently (Andersson \cite{a97}; Kojima \cite{k98}; Beyer and
Kokkotas \cite{bk99}; Kojima and Hosonuma \cite{kh99}).  As in the
newtonian case, a spherical isentropic relativistic star has a large
degenerate subspace of zero-frequency modes consisting of the
axial-parity r-modes \cite{tc67} and the polar-parity g-modes. \cite{t69a}
Although isentropic
newtonian stars retain a vestigial set of purely axial modes (those
having $l=m$), rotating relativistic stars of this type have {\it no}
pure r-modes\cite{alf99}, no modes whose limit for a spherical star is purely
axial.  Instead, the newtonian r-modes with $l=m\geq 2$ acquire
relativistic corrections with both axial and polar parity to become
discrete hybrid modes of the corresponding relativistic models.

In the slow-motion approximation in which they have so far been
studied, $r-modes$ of nonisentropic stars have, remarkably, a {\sl
continuous} spectrum.  Kojima shows that the axial modes are described
by a single, second-order ODE \cite{k98} for the modes' radial
behavior; he argues that the continuous spectrum is implied by the
vanishing of the  coefficient of the highest derivative term of this
equation at some value of the radial coordinate, and Beyer and
Kokkotas\cite{bk99} make the claim precise.  As the latter authors
point out, the continuous spectrum they find may be an artifact of the
vanishing of the imaginary part of the  frequency in the slow rotation
limit.  (Or, more broadly, it may be an artifact of the slow rotation
approximation.)

In addition, Kojima and Hosonuma \cite{kh99} have studied the mixing of
axial and polar perturbations to order $\Omega^2$ in rotating relativistic 
stars, again finding a continuous mode spectrum.  Their calculation uses 
the Cowling approximation (which ignores all metric perturbations) and 
assumes an ordering of the perturbation variables in powers of $\Omega$ 
which forbids the mixing of axial and polar terms at zeroth order. Again
the continuous spectrum may be an artifact of an approximation that
enforces a purely real frequency.
\\ 

Finally, several authors have looked at the 
gravitational-wave driven r-mode instability in white dwarfs. 
Andersson et al\cite{akst98} suggested that the instability
may limit the spin of accreting dwarfs; and Hiscock\cite{h98} found that 
an r-mode instability of dwarfs conforming to the Wagoner scenario --  radiating in gravitational waves the angular momentum it gains in 
accretion -- was potentially detectable by space-based detectors, with a preference for a shorter baseline instrument like LISA.  
Lindblom\cite{l99}, however, finds that the growth time of the 
instability is too slow to allow the amplitude to reach saturation 
in less than $6\times 10^9$ yr; and he finds a very restricted 
range of mass and accretion rates for which the star could be 
hot enough for long enough that the r-mode could grow to a 
dynamically significant amplitude.  Although Lindblom's
analysis allows instability in accreting dwarfs larger than 0.9 $M_\odot$ that are spun up by accretion to about $0.85\Omega_K$, observable
consequences of the instability do not seem likely.

{\sl Summary}

\begin{description}
\item
In a newborn, rapidly rotating neutron star, an r-mode instability 
appears to grow until it is in the nonlinear regime.  It then 
steadily radiates the angular momentum of the star until the 
temperature falls below the superfluid transition temperature of
about $10^9\ K$.  By this time, the star's angular velocity will
have decreased to less than 1/5 the maximum (Keplerian) $\Omega$.

\item
The formation of millisecond pulsars from accretion-induced collapse
of white dwarfs is apparently ruled out for periods less than about
3 ms.

\item
The observation of Marshall et al apparently implies that ordinary supernovae produce neutron stars rotating fast enough to have 
unstable axial modes.

\item  LIGO II may be able to detect the gravitational radiation
emitted by this instability in supernovae within 4 to 20 Mpc. \\ 

\end{description}

All of these conclusions are based on linear perturbation theory with
no magnetic fields, and work is in progress to decide whether the 
maximum mode amplitude is large enough to justify the dramatic 
predictions. \\

\end{document}